\documentclass[11pt,sort&compress]{elsarticle}   	
\usepackage{amssymb}

\usepackage{lineno,hyperref,geometry,amsmath,appendix}
\modulolinenumbers[5]
\usepackage{bbold}
\newcommand{\gdualn}[1]{\overset{\:{}^{{}^{\boldsymbol{\neg}}}}{\smash[t]{#1}}} 
\def\0{\mbox{\boldmath$\displaystyle\mathbb{O}$}}
\def\C{\mbox{\boldmath$\displaystyle\mathbb{C}$}}
\def\R{\mbox{\boldmath$\displaystyle\mathbb{R}$}}

\def\x{\mbox{\boldmath$\displaystyle\boldsymbol{x}$}}

\def\I{\openone}
\def\s{\mbox{\boldmath$\displaystyle\boldsymbol{\sigma}$}}
\def\p{\mbox{\boldmath$\displaystyle\boldsymbol{p}$}}
\def\0{\mbox{\boldmath$\displaystyle\boldsymbol{0}$}}

\def\openone{\mathbb I}
\def\pii{\mbox{\boldmath$\displaystyle\boldsymbol{\pi}$}}

\journal{Proceedings of the Royal Society A}

\begin{document}

\begin{frontmatter}
\title{A new class of mass dimension one fermions}

\author[mymainaddress]{Dharam Vir Ahluwalia\corref{mycorrespondingauthor}}
\cortext[mycorrespondingauthor]{Corresponding author}
\ead{dharam.v.ahluwalia@gmail.com}

\address[mymainaddress]{Mountain Physics Camp, Center for the Studies of the Glass Bead Game\\  Bir, Himachal Pradesh, 176077
India}


\begin{abstract}
These are notes on the square root of $4\times4$ identity matrix and  associated quantum fields of spin one half. The method is illustrated by constructing a new mass dimension one fermionic field. The presented field is local. The field energy is bounded from below. It is argued that these  fermions are a first-principle candidate for dark matter with an
unsuppressed quartic self interaction. \end{abstract}

\end{frontmatter}
\vspace{21pt}

\noindent
Historically, Dirac equation arose in taking the square root of the dispersion relation
$
p^\mu p_\mu = m^2$~\cite{Dirac:1928hu}
The square root of the left hand side was found to be $\gamma_\mu p^\mu$, where the $\gamma_\mu$ are the celebrated $4\times 4$ matrices of the Dirac framework. The argument naturally leads to 
$(\gamma_\mu p^\mu \pm m \I)\psi(\p) =0$, the Dirac equation in momentum space.
Its solutions, after attending to certain locality phases, later became expansion coefficients of all fermionic matter fields of the standard model~\cite{Weinberg:1995mt}. 
\vspace{5pt}

Modulo the Majorana observation of 1937~\cite{Majorana:1937vz}, there is a general consensus that the Dirac field presents a unique spin one half field that is consistent with Lorentz symmetries and locality.
The uniqueness, however, hinges on the implicit assumption that the square root of a $4\times4$ identity matrix $\I$ multiplying the $m^2$  on the right hand side, is $\I$ itself.  The recent emergence of the new spin one-half fermions with mass dimension one provides a strong reason that other roots of $\I$ may lead to new spin one half matter fields, and these may serve the dark matter sector or at the least  provide us with a complete set of particle content consistent with basic principles of quantum mechanics and symmetries of special 
relativity~\cite{Ahluwalia:2019etz,Ahluwalia:2016rwl,Ahluwalia:2015vea,Pereira:2016eez,Bueno_Rogerio_2018,Pereira_2019,Hoff_da_Silva_2019,Bueno_Rogerio_2019,Sorkhi:2002,Zhou_2018,KaiErikWunderle:2010,Jardim:2014xla,Boehmer:2010ma,Boehmer:2006qq,Bahamonde_2018,Basak:2014qea}.
\vspace{5pt}

With this background and motivation we recall the well known linearly independent square roots of identity~\cite[p. 71]{Schweber:1961zz}
\begin{align}
&\I\\
&i\gamma_1 \quad  i \gamma_2\quad i\gamma_3 \quad\gamma_0\\
& i \gamma_2\gamma_3\quad i\gamma_3\gamma_1 \quad i \gamma_1\gamma_2\quad\gamma_0\gamma_1\quad\gamma_0\gamma_2\quad\gamma_0
\gamma_3\\
&i\gamma_0\gamma_2\gamma_3\quad
i\gamma_0\gamma_1\gamma_3\quad
i\gamma_0\gamma_1\gamma_2\quad
\gamma_1\gamma_2\gamma_3\\
&i\gamma_0\gamma_1\gamma_2\gamma_3
\end{align}
We denote these by  $\Gamma_\ell$, $\ell=1,2\cdots 16$, with $\Gamma_1$ being the first entry in the above array and $\Gamma_{16}$ being the last -- $\ell$ assignment is in consecutive order. We shall adopt the Weyl basis in the $(1/2,0)\oplus(0,1/2)$ representations space. 

\vspace{5pt}

\noindent
To illustrate the method we consider $\Gamma_7$.
Its eigenspinors, up to constant multiplicative factors, are
\begin{equation}
\lambda_1 = \left(
\begin{array}{c}
 0 \\
 0 \\
 -i \\
 1 \\
\end{array}
\right),\quad
\lambda_2 =  \left(
\begin{array}{c}
 0 \\
 0 \\
 i \\
 1 \\
\end{array}
\right),\quad
\lambda_3=\left(
\begin{array}{c}
 -i \\
 1 \\
 0 \\
 0 \\
\end{array}
\right),\quad
\lambda_4=\left(
\begin{array}{c}
 i \\
 1 \\
 0 \\
 0 \\
\end{array}
\right).\label{eq:lambda}
\end{equation}
The first and the third eigenspinors correspond to eigenvalue $+1$ of $\Gamma_7$, and the other  two to eigenvalue  $-1$ of $\Gamma_7$.
We define these as the `rest spinors' $\lambda_i(0)$. 

By acting the boost operator
\begin{equation}
\kappa = \sqrt{\frac{E+m}{2 m}}\left[\begin{array}{cc} 
\I + \frac{\s\cdot\p}{E+m} & \0 \\
\0 &  \I- \frac{\s\cdot\p}{E+m}
\end{array}\right]
\end{equation}
on these spinors we obtain the four eigenspinors for an arbitrary momentum $\lambda_i(\p) = \kappa\lambda_i(0)$. We implement our programme by solving the following four 
equations for $\tau_{ij}\in \R$:
\begin{align}
&m^{-1} \gamma_\mu p^\mu \lambda_1(\p) 
- \tau_{13}  \lambda_3(\p) =0,\quad
m^{-1} \gamma_\mu p^\mu \lambda_2(\p) 
- \tau_{24}  \lambda_4(\p) =0\\
& m^{-1} \gamma_\mu p^\mu \lambda_3(\p) 
- \tau_{31}  \lambda_1(\p) =0,\quad
m^{-1} \gamma_\mu p^\mu \lambda_4(\p) 
- \tau_{42}  \lambda_2(\p) =0
\end{align}
and find that a single $\tau$, equal to unity, solves all the four equations and assures that while $\lambda_i(\p)$ do not  satisfy the Dirac equation they instead satisfy the spinorial Klein-Gordon equation. We thus pass the first test for the viability of the theory to be Lorentz covariant. 
\vspace{5pt}

To study the CPT properties of $\lambda(\p)$ we 
introduce $\Theta$, the Wigner time reversal operator, and $\gamma$ 
\begin{align}
&\Theta =\left(\begin{array}{cc}
0 &-1\\
1 &0
\end{array}
\right),\quad
 \gamma = \frac{i}{4!} \epsilon_{\mu\nu\lambda\sigma}
\gamma^\mu\gamma^\nu\gamma^\lambda\gamma^\sigma = \left(\begin{array}{cc}
\I &\0\\
\0 & -\I
\end{array}\right)\label{eq:gamma5}
\end{align}
where $\epsilon_{\mu\nu\lambda\sigma}$ is the
completely antisymmetric 4th rank tensor with $\epsilon_{0123} = +1$ (the dimensionality of identity matrix $\I$ and null matrix $\0$ shall be apparent from the context)].
The charge conjugation $\mathcal{C}$, parity $\mathcal{P}$, and time reversal $\mathcal{T}$, operators can then be written as
\begin{equation}
\mathcal{C} = \left(\begin{array}{cc}
\0 & i\Theta \\
-i\Theta &\0
\end{array}
\right) K,\quad
\mathcal{P}= m^{-1} \gamma_\mu p^\mu,\quad
\mathcal{T} = i \gamma \mathcal{C}
\end{equation}
where $K$ complex conjugates to the right. We then readily obtain
\begin{align}
&\mathcal{C} \lambda_1(\p) = - \lambda_4(\p),\quad
\mathcal{C} \lambda_2(\p) =  \lambda_3(\p),\quad
\mathcal{C} \lambda_3(\p) =  \lambda_2(\p),\quad
\mathcal{C} \lambda_4(\p) = - \lambda_1(\p), \label{eq:C}\\
&\mathcal{P} \lambda_1(\p) = \lambda_3(\p),\quad
\mathcal{P} \lambda_2(\p) =  \lambda_4(\p),\quad
\mathcal{P} \lambda_3(\p) =  \lambda_1(\p),\quad
\mathcal{P} \lambda_4(\p) =  \lambda_2(\p), \label{eq:P}\\
&\mathcal{T} \lambda_1(\p) = - i \lambda_4(\p),\quad
\mathcal{T} \lambda_2(\p) =  i \lambda_3(\p),\quad
\mathcal{T} \lambda_3(\p) =  -i \lambda_2(\p),\quad
\mathcal{T} \lambda_4(\p) = i \lambda_1(\p)\label{eq:T}
\end{align}
with the consequence that $(\mathcal{CPT})^2=\I$, with $\mathcal{C}^2=\I$, $ \mathcal{P}^2=\I$, $\mathcal{T}^2=-\I$. The charge conjugation and parity operators anticommute:
$
\{\mathcal{C},\mathcal{P}\} = 0.
$

\vspace{5pt}
As in the case for Elko~\cite{Ahluwalia:2019etz,lee2019spinhalf}, here too we find that under the Dirac dual each of the  $\lambda_i(\p)$, $i=1,2,3,4$,  has null norm. As such we define a new dual:\footnote{The freedom in the definition of spinorial duals was first pointed out in an unpublished e-print of the author~\cite{Ahluwalia:2003jt}, and after several intervening publications it takes its final form for Elko in ~\cite{Ahluwalia:2016rwl}. The subject
has now developed into a research sub-field of its own. We refer the reader to~\cite{Cavalcanti_2020} for a sense of excitement and relevant references.} 
\begin{align}
&  \gdualn{\lambda}_1(\p)  =\big[+ \mathcal{P} \lambda_1(\p)\big]^\dagger\gamma_0 = \overline{\lambda}_3(\p),\quad
\gdualn{\lambda}_2(\p) = \big[+ \mathcal{P} \lambda_2(\p)\big]^\dagger\gamma_0  = \overline{\lambda}_4(\p),\\
 & \gdualn{\lambda}_3(\p)  = \big[- \mathcal{P} \lambda_3(\p)\big]^\dagger\gamma_0  = - \overline{\lambda}_1(\p),\quad
\gdualn{\lambda}_4(\p) = \big[- \mathcal{P} \lambda_4(\p)\big]^\dagger\gamma_0  =  -\overline{\lambda}_2(\p).\label{eq:pod}
\end{align}
 After re-norming the rest eigenspinors by a multiplicative factor of $\sqrt{m}$,  the new dual gives the following Lorentz invariant orthonormality relations 
 \begin{align}
&\gdualn{\lambda}_i(\p) {\lambda}_i(\p) = +2m,\quad i=1,2\label{eq:on12}\\
&\gdualn{\lambda}_i(\p) {\lambda}_i(\p) = - 2m,\quad i=3,4\label{eq:on34}
\end{align}
with cross terms identically zero, and the `spin sums'
\begin{align}
\sum_{i=1,2}{\lambda}_i(\p) \gdualn{\lambda}_i(\p)  = 2 m\left(
\begin{array}{cccc}
0 & 0 & 0 & 0\\
0 & 0 & 0 & 0\\
0 & 0 & 1 & 0\\
0 & 0 & 0 &1
\end{array}\right),\quad
\sum_{i=3,4}{\lambda}_i(\p) \gdualn{\lambda}_i(\p)  = -  2 m\left(
\begin{array}{cccc}
1 & 0 & 0 & 0\\
0 & 1 & 0 & 0\\
0 & 0 & 0 & 0\\
0 & 0 & 0 &0
\end{array}\right) \label{eq:ss}
\end{align}
leading to the completeness relation
\begin{equation}
\frac{1}{2 m}
\left[\sum_{i=1,2} {\lambda}_i(\p) \gdualn{\lambda}_i(\p) - 
\sum_{i=3,4} {\lambda}_i(\p) \gdualn{\lambda}_i(\p) \right]= \I .\label{eq:spinsums}
\end{equation}

We thus introduce a new spin one half quantum field with  $\lambda_i(\p)$ as its expansion co-efficients:
\begin{equation}
\mathfrak{b}(x) \stackrel{\mathrm{def}}{=}
\int\frac{\mbox{d}^3 p}{(2\pi)^3}
\frac{1}{\sqrt{2 m E(\p)}}
\bigg[
\sum_{i=1,2} {a}_i(\p)\lambda_i(\p) e^{-i p\cdot x}+ \sum_{i=3,4} b^\dagger_i(\p)\lambda_i(\p) e^{i p\cdot x}\bigg]\label{eq:fieldb}
\end{equation}
with 
\begin{equation}
\gdualn{\mathfrak{b}}(x) \stackrel{\mathrm{def}}{=}
\int\frac{\mbox{d}^3 p}{(2\pi)^3}
\frac{1}{\sqrt{2 m E(\p)}}
\bigg[
\sum_{i=1,2} a^\dagger_i(\p)\gdualn{\lambda}_i(\p) e^{i p\cdot x}+ \sum_{i=3,4} b_i(\p)\gdualn{\lambda}_i(\p) e^{-i p\cdot x}\bigg]
\end{equation}
as its adjoint.
At this stage we do not fix the statistics to be fermionic 
\begin{equation}
\left\{a_i(\p),a^\dagger_j(\p)\right\} =(2 \pi)^3 \delta^3(\p-\p^\prime)\delta_{ij}, \quad \left\{a_i(\p), a_j(\p^\prime)\right\} = 0 =
 \left\{a^\dagger_i(\p), a^\dagger_j(\p^\prime)\right\}\label{eq:b}
\end{equation}
or bosonic
\begin{equation}
\left[a_i(\p),a^\dagger_j(\p)\right] =(2 \pi)^3 \delta^3(\p-\p^\prime)\delta_{ij}, \quad \left[a_i(\p), a_j(\p^\prime)\right] = 0 =
 \left[a^\dagger_i(\p), a^\dagger_j(\p^\prime)\right]\label{eq:bd}
\end{equation}
and assume similar anti-commutation, or commutation, relations for $b_i(\p)$ and $b_i^\dagger(\p)$. 
\vspace{5pt}

 To determine the statistics for the 
$\mathfrak{b}(x)$ and $\gdualn{b}(x)$ system we consider two events, $x$ and $x'$, and note that  the amplitude to propagate from $x$ to $x'$ is then
\begin{align}
\mathcal{A}_{x\to x^\prime}   =  \xi \Big(\underbrace{\langle\hspace{3pt}\vert
\mathfrak{b}(x^\prime)\gdualn{\mathfrak{b}}(x)\vert\hspace{3pt}\rangle \theta(t^\prime-t)
\pm  \langle\hspace{3pt}\vert
\gdualn{\mathfrak{b}}(x) \mathfrak{b}(x^\prime)\vert\hspace{3pt}\rangle \theta(t-t^\prime)}_{\langle\hspace{4pt}\vert \mathfrak{T} ( \mathfrak{b}(x^\prime) \gdualn{\mathfrak{b}}(x)\vert\hspace{4pt}\rangle}\Big)\label{eq:Axtoxprime}
\end{align}
where 
\begin{itemize}
 \item[\textemdash] the plus sign holds for bosons and the minus sign for fermions,

\item[\textemdash] $\xi\in\C$ is to be determined from the normalisation condition that
$
 \mathcal{A}_{x\to x^\prime} 
$
integrated over all possible separations $x-x^\prime$
 be unity (or, more precisely  $e^{i\gamma}$, with $\gamma\in \R$). 
\item[\textemdash] 
and $\mathfrak{T}$ is the time ordering operator.
\end{itemize}
The two vacuum expectation values that appear in $\mathcal{A}_{x\to x^\prime} $ evaluate to the following expressions
 \begin{align}
\langle\hspace{3pt}\vert
\mathfrak{b}(x^\prime)\gdualn{\mathfrak{b}}(x)\vert\hspace{3pt}\rangle  & =\int\frac{\text{d}^3p}{(2 \pi)^3}\left(\frac{1}{2 m E(\p)}\right)
 e^{-ip\cdot(x^\prime-x)}
 \sum_{i=1,2}\lambda_i(\p)
\gdualn\lambda_i(\p) \label{eq:amplitudeP-newS}
\\
\langle\hspace{3pt}\vert
\gdualn{\mathfrak{b}}(x) \mathfrak{b}(x^\prime)\vert\hspace{3pt}\rangle 
  & =  \int\frac{\text{d}^3p}{(2 \pi)^3}\left(\frac{1}{2 m E(\p)}\right)
 e^{ip\cdot(x^\prime-x)}
 \sum_{i=3,4}\lambda_i(\p)
\gdualn\lambda_i(\p) .\label{eq:amplitudeP-newA}
\end{align}
The two Heaviside step functions of equation
(\ref{eq:Axtoxprime}) can now be replaced by their integral representations
\begin{align}
\theta(t^\prime-t) &= \lim_{\epsilon\to 0^+} \int\frac{\text{d}\omega}{2\pi i}
\frac{e^{i \omega (t^\prime-t)}}{\omega- i \epsilon} \\
\theta(t-t^\prime) &= \lim_{\epsilon\to 0^+} \int\frac{\text{d}\omega}{2\pi i}
\frac{e^{i \omega (t-t^\prime)}}{\omega- i \epsilon}
\end{align}
where $\epsilon,\omega\in\R$. Using these results, and  
\begin{itemize}
\item  substituting $\omega \to p_0 = -\omega+E(\p)$ in the first term and  $\omega \to p_0 = \omega- E(\p)$ in the second term
\item and using the results (\ref{eq:spinsums})
for the spin sums
\end{itemize}
we are forced -- by internal consistency of the resulting formalism -- to pick the minus sign in (\ref{eq:Axtoxprime}), giving 
\begin{equation}
\mathcal{A}_{x\to x^\prime}  = i \,2 \xi \int\frac{\text{d}^4 p}{(2 \pi)^4}\,
e^{-i p_\mu(x^{\prime\mu}-x^\mu)}
\frac{\I}{p_\mu p^\mu -m^2 + i\epsilon}\label{eq:AmplitudeWithXi}
\end{equation}
This is equivalent to the choice   (\ref{eq:b}) over (\ref{eq:bd}). Following~ \cite{Ahluwalia:2019etz}, the normalisation $\xi$ is seen to be \cite{Ahluwalia:2019etz}
\begin{equation}
\xi = \frac{i m^2}{2}
\end{equation}
resulting in
\begin{equation}
\mathcal{A}_{x\to x^\prime}  = - m^2 \int\frac{\text{d}^4 p}{(2 \pi)^4}\,
e^{-i p_\mu(x^{\prime\mu}-x^\mu)}
\frac{\I}{p_\mu p^\mu -m^2 + i\epsilon}
\end{equation}
We define the Feynman-Dyson propagator 
\begin{align}
S_{\textrm{FD}}(x^\prime-x) & \stackrel{\textrm{def}}{=} - \frac{1}{m^2} 
\mathcal{A}_{x\to x^\prime}\nonumber\\
 &= \int\frac{\text{d}^4 p}{(2 \pi)^4}\,
e^{-i p_\mu(x^{\prime\mu}-x^\mu)}
\frac{\I_4}{p_\mu p^\mu -m^2 + i\epsilon}\label{eq:FD-prop-b}
\end{align}
so that
\begin{equation}
\left(\partial_{\mu^\prime} \partial^{\mu^\prime} \I_4 + m^2\I_4\right)
S_{\textrm{FD}}(x^\prime-x)  = -  \delta^4(x^\prime - x)\label{eq:KGDiracDelta}
\end{equation}
In terms of the new field $\mathfrak{b}(x)$ and its adjoint $\gdualn{\mathfrak{b}}(x)$ it takes the form
\begin{equation}
S_{\textrm{FD}}(x^\prime-x)= -\frac{i}{2}\left\langle\hspace{4pt}\left\vert \mathfrak{T} ( \mathfrak{b}(x^\prime) \gdualn{b}(x)\right\vert\hspace{4pt}\right\rangle\label{eq:propagator}
\end{equation}
and establishes the mass dimension of the field to be one, leading to the following free field Lagrangian density
\begin{equation}
\mathfrak{L}(x) =\partial^\mu\gdualn{\mathfrak{b}}\,\partial_\mu {\mathfrak{b}}(x) - m^2 \gdualn{\mathfrak{b}}(x) \mathfrak{b}(x)\label{eq:fieldlagrangian}
\end{equation}

This determines the momentum conjugate to $\mathfrak{b}(x)$
\begin{equation}
\pii(x) = \frac{\partial \mathfrak{L}(x)}
{\partial {\dot{\mathfrak{b}}(x)}} = \frac{\partial}{\partial t}\gdualn{\mathfrak{b}}(x).\label{eq:px}
\end{equation}
Using the spin sums given in equation (\ref{eq:ss}) we determine the locality structure of the new fermionic field to be
\begin{align}
&\left\{\mathfrak{b}(t,\x),\;\pii(t,\x^\prime) \right\} = i \delta^3\left(\x-\x^\prime\right)\I,\quad\label{eq:lac-2and3c}\\
&\left\{ \mathfrak{b}(t,\x),\;\mathfrak{b}(t,\x^\prime) \right\}= 0, \quad 
\left\{ \pii(t,\x),\;\pi(t,\x^\prime) \right\} = 0.\label{eq:lac-2and3}
\end{align}
To examine if the energy associated with the introduced $\mathfrak{b}(x)$-$\gdualn{\mathfrak{b}}(x)$ system has  the usual zero point contribution and is bounded from below, we carry out a calculation similar to the one presented in \cite[Section 7]{Ahluwalia:2004ab} and find the field energy to be
\begin{equation}
H = \int\frac{\text{d}^3 p}{(2\pi)^3}\frac{1}{2 m} E(\p)\Bigg[\sum_{i=1,2} a_i^\dagger(\p) a_i(\p) \gdualn{\lambda}_i(\p) \lambda_i(\p) 
+
\sum_{i=3,4} b_i(\p) b_i^\dagger(\p) \gdualn{\lambda}_i(\p) \lambda_i(\p)
\Bigg]
\end{equation}
Use of the orthonormality relations~(\ref{eq:on12}) and~(\ref{eq:on34}) reduce the above expression to
\begin{equation}
H = \int\frac{\text{d}^3 p}{(2\pi)^3} E(\p)\Bigg[\sum_{i=1,2} a_i^\dagger(\p) a_i(\p) 
-
\sum_{i=3,4} b_i(\p) b^\dagger_i(\p)
\Bigg]
\end{equation}
Consistent with the obtained fermionic locality anticommutator~(\ref{eq:lac-2and3c}), the next simplification occurs by exploiting
\begin{equation}
\{ b_i(\p),b_{i^\prime}(\p^\prime)\}=(2\pi)^3\delta^3(\p-\p^\prime)\delta_{i i^\prime}
\label{eq:stat}
\end{equation}
with the result that
\begin{equation}
H= \underbrace{-\,\delta^3(\0) \int\text{d}^3p \; 2 E(\p) }_{H_0}\;+ \sum_{i=1,2} \int\frac{\text{d}^3 p}{(2 \pi)^3}
E(\p) a_i^\dagger(\p) a_i(\p)
+ \sum_{i=3,4} \int\frac{\text{d}^3 p}{(2 \pi)^3}
E(\p) b_i^\dagger(\p) b_i(\p)\nonumber
\end{equation}
To obtain a representation for $\delta^3(\0) $ that appears in the above expression for the field energy, we note that since $\delta^3(\p)$ may be expanded as 
\begin{equation}
\delta^3(\p) = \frac{1}{(2\pi)^3}\int\text{d}^3 x \exp(i \p\cdot \x)
\end{equation}
 $\delta^3(\0)$ may be replaced by  $[1/(2\pi)^3]\int\text{d}^3 x$, giving the following contribution for the zero point energy
 \begin{equation}
 H_0 = - 4 \times \frac{1}{(2\pi)^3}\int\text{d}^3 x\int\text{d}^3 p \; \frac{1}{2} E(\p)
 \end{equation}
 Since in natural units $\hbar$ is set to unity, $ \frac{1}{(2\pi)^3} \text{d}^3 x \,\text{d}^3 p $ acquires the interpretation of a unit-size phase cell,  with $-\frac{1}{2} E(p)$ as its energy content.
The factor of $4$ in the expression for $H_0$ corresponds to the four particle and antiparticle degrees carried
by the $\mathfrak{b}(x)$-$\gdualn{\mathfrak{b}}(x)$ system. The remaining two terms in the expression for $H$ establish that for a given momentum $\p$ each of the four particle-antiparticle degrees of freedom contributes equally.
\vspace{21pt}

This completes our construction of an entirely new class of spin one half fermions. Their physical implications are essentially unknown. Because of the mass dimensionality mismatch with the standard model fermions -- $3/2$ versus $1$ -- the new fermions cannot enter the standard model doublets. For this reason they  are a natural dark matter candidate with unsuppressed quartic self interaction.
\vspace{21pt}

\noindent
\textbf{Note.} Because the results presented here have evolved out of a manuscript arXived as e-print~\cite{Ahluwalia:2019ujt} it is important to make a remark.
The point of departure starts with equation~(\ref{eq:pod}), that is the definitions of the duals  of $\lambda_3(\p)$ and $\lambda_4(\p)$. This change percolates through the rest of the calculations, finally replacing 
\begin{align}
&\left[\mathfrak{b}(t,\x),\;\mathfrak{p}(t,\x^\prime) \right] = i \delta^3\left(\x-\x^\prime\right) \openone_\ell,\quad\nonumber\\
&\left[ \mathfrak{b}(t,\x),\;\mathfrak{b}(t,\x^\prime) \right]= 0, 
\left[ \mathfrak{p}(t,\x),\;\mathfrak{p}(t,\x^\prime) \right] = 0.\nonumber
\end{align}
where
\begin{equation}
\openone_\ell \stackrel{\mathrm{def}}{=} \left(
\begin{array}{cccc}
-1 & 0 &  0 & 0\\
0 & -1 & 0 & 0\\
0 & 0 & 1 & 0\\
0 & 0& 0& 1
\end{array}\right)\nonumber
\end{equation}
of the e-print by  equations~(\ref{eq:lac-2and3c}) and (\ref{eq:lac-2and3}). This noted, if one proceeds with commutator counterparts of~(\ref{eq:stat}), and keeps the duals of $\lambda_3(\p)$ and $\lambda_4(\p)$ as before, the resulting field energy is found to be same as above but with $H_0$ replaced by 
\begin{equation}
 H_0 = + 4 \; \times \frac{1}{(2\pi)^3}\int\text{d}^3 x\int\text{d}^3 p \; \frac{1}{2} E(\p).\nonumber
 \end{equation}
\vspace{21pt}

\noindent
\textbf{Funding.} The research presented here is entirely supported by the personal funds of the author.
\vspace{5pt}

\noindent
\textbf{Acknowledgements.} I am grateful to the two anonymous referees who reviewed and commented constructively. I thank  
Julio Marny Hoff da Silva and Cheng-Yang Lee for correspondence related to the ideas presented here, and Sweta Sarmah for discussions.

\vspace{21pt}

\noindent
\textbf{References}

\end{document}